\documentclass[fleqn,usenatbib]{mnras}

\usepackage{newtxtext,newtxmath}
\usepackage[T1]{fontenc}

\DeclareRobustCommand{\VAN}[3]{#2}
\let\VANthebibliography\thebibliography
\def\thebibliography{\DeclareRobustCommand{\VAN}[3]{##3}\VANthebibliography}

\usepackage{graphicx}	% Including figure files
\usepackage{amsmath}	% Advanced maths commands
\usepackage{amssymb}	% Extra maths symbols
\usepackage{booktabs}   %Will addition
\usepackage{multirow}   %will addition
\usepackage{comment}    %Will addition
\usepackage[utf8]{inputenc} %will addition
\title[Stellar mass drives the MZR]{Stellar mass, not dynamical mass nor gravitational potential, drives the mass-metallicity relationship}
\author[W. M. Baker et al.]
{William M. Baker$^{1,2}$\thanks{E-mail: wb308@cam.ac.uk},
Roberto Maiolino$^{1,2,3}$\\
$^{1}$Kavli Institute for Cosmology, University of Cambridge, Madingley Road, Cambridge, CB3 OHA, UK\\
$^{2}$Cavendish Laboratory - Astrophysics Group, University of Cambridge, 19 JJ Thomson Avenue, Cambridge, CB3 OHE, UK\\
$^{3}$Department of Physics and Astronomy, University College London, Gower Street, London WC1E 6BT, UK\\
}

\date{Accepted XXX. Received YYY; in original form ZZZ}

\pubyear{2021}

\begin{document}
\label{firstpage}
\pagerange{\pageref{firstpage}--\pageref{lastpage}}
\maketitle

% Abstract of the paper
\begin{abstract}
The widely known relation between stellar mass and gas metallicity (mass-metallicity relation, MZR) in galaxies is often ascribed to the higher capability of more massive systems to retain metals against the action of galactic outflows. In this scenario the stellar mass would simply be an indirect proxy of the dynamical mass or of the gravitational potential.
We test this scenario by using a sample of more than one thousand star-forming galaxies from the MaNGA survey for which dynamical masses have been accurately determined. By using three different methods (average dispersion, Partial Correlation Coefficients, Random Forest) we unambiguously find that the gas metallicity depends primarily and fundamentally on the stellar mass. Once the dependence on stellar mass is taken into account, there is little or no dependence on either dynamical mass or gravitational potential (and, if anything, the metallicity dependence on the latter quantities is inverted). Our result indicates that the MZR is not caused by the retention of metals in more massive galaxies. The direct, fundamental dependence of metallicity on stellar mass suggests the much simpler scenario in which the MZR is just a consequence of the stellar mass being proportional to the integral of metals production in the galaxy.
\end{abstract}

% Select between one and six entries from the list of approved keywords.
% Don't make up new ones.

%%%%%%%%%%%%%%%%%%%%%%%%%%%%%%%%%%%%%%%%%%%%%%%%%%

%%%%%%%%%%%%%%%%% BODY OF PAPER %%%%%%%%%%%%%%%%%%

\section{Introduction}

The mass-metallicity relation \citep[MZR,][e.g. ]{2004Tremonti, Maiolino2019} describes how the metallicity of galaxies is linked to their stellar mass. The greater the stellar mass of a galaxy, the higher its metallicity, although the metallicity plateaus at high stellar masses \citep[log($\rm M_*/M_\odot)>$ 10.5 - 11.0, ][]{2004Tremonti}.
The MZR applies for both the gas-phase and stellar metallicities \citep[][]{Trager_Z_2000AJ....120..165T,Zahid_2017ApJ...847...18Z}, but here we focus on the gas-phase metallicities only, and for star-forming galaxies.

The mass-metallicity relation has also been found to hold on resolved ($\sim$kpc) scales \citep[][]{RosalesOrtega2012ApJ...756L..31R} in what is known as the resolved mass metallicity relation \citep[rMZR,][]{Barrera_global_from_local_radial2016MNRAS.463.2513B,Almeida_and_Menguiano2019ApJ...878L...6S, Baker_rFMR_2022arXiv221003755B}.

The origin of the MZR has been highly debated \citep[see][for a review]{Maiolino2019}. 

A common interpretation is that the deeper gravitational potential in massive galaxies is capable of better retaining metals against outflows \citep[][]{2004Tremonti, Tumlinson_2011Sci...334..948T, Chisholm_outflows_2018MNRAS.481.1690C}.
 This scenario requires galactic outflows with a mass-dependent loading factor (i.e. decreasing at high masses). It also requires that outflows are more metal loaded than the interstellar medium (ISM).
 Otherwise a generic outflow of material with a metal enrichment equal to the ISM could not decrease the metallicity of the ISM. Outflows with a greater metallicity than the parent galaxy's ISM have indeed been observed \citep{Chisholm_outflows_2018MNRAS.481.1690C, Konami_metal_loaded_outflow2011PASJ...63S.913K, Origlia_metal_loaded_winds2004ApJ...606..862O}.

If this is the explanation for the MZR, then the correlation of metallicity with stellar mass is because M$_*$ is simply a proxy of the dynamical mass ($\rm M_{dyn}$) and of the gravitational potential ($\phi$). So if we could directly measure $\rm M_{dyn}$ or $\phi$, a tighter correction with metallicity would be expected.

Within this context, recent work has also highlighted the importance of $\Phi_*=M_*/R_e$ as a proxy of the gravitational potential \citep[][]{D'Eugenio_potential_2018MNRAS.479.1807D,Cappelari_M/R_2022arXiv220814974C}.
\citet{D'Eugenio_potential_2018MNRAS.479.1807D} found that gas-phase metallicity correlates more tightly with $\Phi_*$ than M$_*$, which they attribute to the explanation of the gravitational potential being better able to retain metals. This was also found for stellar metallicities \citep{Barone2020ApJ...898...62B, Cappelari_M/R_2022arXiv220814974C, Vaughan_Z_phi_2022MNRAS.tmp.2242V}. In addition to the previously mentioned quantities the velocity dispersion ($\sigma$) has also been suggested as an important parameter driving the metallicity \citep[][]{Li_dynamicalMass_2018MNRAS.476.1765L,Cappelari_M/R_2022arXiv220814974C}.

 The mass-metallicity relation is also observed to evolve with redshift \citep[][]{Maiolino2008A&A...488..463M,Troncoso2014A&A...563A..58T, Sanders2021ApJ...914...19S, Curti_JWST_2022arXiv220712375C}, where this observed evolution can be understood to arise due to the MZR tracing the evolution of the star formation rate (SFR) with redshift as part of the Fundamental Metallicity Relation \citep[FMR,][]{2008Ellison,2010Mannucci, 2020CurtiMNRAS.491..944C, Baker_rFMR_2022arXiv221003755B}.

However, directly testing the correlation of metallicity with $\rm M_{dyn}$ and $\phi$ is feasible now thanks to the MaNGA \citep[Mapping Nearby Galaxies at Apache Point Observatory][]{Bundy2015} survey, which provides integral field spectroscopy (hence 2D kinematic information) for thousands of galaxies.

In this paper
we use local star-forming galaxies from MaNGA to explore the relative importances of stellar mass, dynamical mass and $\phi$ in driving the gas-phase metallicity using partial correlation coefficients and random forest regression. This enables us to directly test whether the aforementioned explanation of the MZR is plausible.

In Section \ref{sec:Data_and_methods} we introduce the MaNGA survey. In Section \ref{sec:methods} we introduce the derivation of the physical quantities and in Section \ref{sec:disp_pcc_rf} we introduce our analysis techniques. In Section \ref{sec:results} we explore the results of this analysis and in Section \ref{sec:discussion} we discuss their significance. Section \ref{sec:Conclusions} provides a short summary of the key results.

In this paper we assume that  $\rm H_0=70$km $\text{s}^{-1}$ $\text{Mpc}^{-1}$, $\Omega_m=0.3$ and $\Omega_{\Lambda}=0.7$.

\section{Data}

\label{sec:Data_and_methods}

%\subsection{MaNGA}
We use publicly available integral field spectroscopic (IFS) data from the MaNGA \citep[Mapping Nearby Galaxies at Apache Point Observatory][]{Bundy2015} survey, conducted during the fourth generation of the Sloan Digital Sky Survey \citep[SDSS][]{Blanton2017AJ....154...28B}. The MaNGA survey observed approximately 10,000 local galaxies, with data sampling out to out to 1.5$R_e$ (effective radii) in 2/3 of the sample and 2.5$R_e$ in the other third. For further details about the MaNGA survey see \citet{Wake2017AJ....154...86W, Law2016AJ....152...83L, Yan2016AJ....151....8Y, Yan2016AJ....152..197Y}.

Out of the whole sample we use accurate dynamical masses provided by \cite{Li_dynamicalMass_2018MNRAS.476.1765L} which were obtained via detailed Jeans Anisotropic modelling \citep[][]{Cappellari_JAM_2008MNRAS.390...71C} for 1158 spiral galaxies in MaNGA. We also use the provided circularized effective radius, $R_e$ which was obtained via  fitting the SDSS r-band image, and the velocity dispersion, $\sigma_e$, obtained within an elliptical aperture of area $A=\pi R_e^2$.

\section{Derivation of physical quantities}

\label{sec:methods}

We use metallicities, stellar masses and star formation rates obtained following a similar method to  \cite{Baker_rFMR_2022arXiv221003755B}. A brief summary is as follows.
The stellar mass surface densities are obtained from PIPE3D \citep[][]{SanchezPipe3d2016RMxAA..52..171S}. 
To obtain the total stellar mass we integrate the stellar mass surface densities of all spaxels within the effective radius. 
The emission line flux maps are obtained from the Data Analysis Pipeline \citep[DAP,][]{BelfioreDAP2019AJ....158..160B, Westfall2019AJ....158..231W}. A S/N cut of 3 is applied to all emission lines (H$\alpha$, H$\beta$, [OII]$\lambda$3727,29, [OIII]$\lambda$5007, [NII]$\lambda$6584, and [SII]$\lambda$6718,32) and an equivalent width cut of 6\AA\ is applied to H$\alpha$ to minimise contamination from diffuse ionised gas (which would affect the validity of the metallicity calibrations).
The emission lines are extinction corrected using the H$\alpha$/H$\beta$ Balmer decrement (assuming an intinsic value of 2.86) with a \citet{Calzetti2000} attenuation curve. As the metallicity diagnostics are only calibrated for star-forming regions selected as such in the [NII]-BPT diagram, we select them via the [NII]-BPT \citep[][]{Baldwin} diagram with the \citet{2003Kauffmann} diagnostic curve.
The gas-phase metallicities are obtained for the resulting star-forming regions using a combination of nine metallicity diagnostics with the  calibrations detailed in \citet{2017MNRASCurti,2020CurtiMNRAS.491..944C}. In order to obtain a single metallicity for each galaxy, we take the average metallicity of the star forming regions within the effective radius of each galaxy. We will also investigate the dependence on star formation rate (SFR). The SFR surface density for each SF spaxel is obtained from the H$\alpha$ emission line following the conversion detailed in \citet{Kennicutt2012ARA&A..50..531K}. The total star formation rate is then calculated by integrating the individual star formation rate surface densities of all star-forming spaxels within the effective radius of the galaxy, i.e. the same region within which we calculate the metallicity. 
In the exploration of some relations we remove galaxies that are outliers by sigma-clipping from the median of the specific relation, with the cutoff being 5$\sigma$. 

As mentioned in Section \ref{sec:Data_and_methods}, we also use dynamical masses, circularized effective radii and velocity dispersions from \cite{Li_dynamicalMass_2018MNRAS.476.1765L}. These were obtained via Jeans Anisotropic Modelling \citep[JAM][]{Cappellari_JAM_2008MNRAS.390...71C}. In particular, the dynamical mass was defined from the best-fitting JAM models as the total mass enclosed within a spherical half-light radius. Essentially, this means that the dynamical masses of the galaxies are primarily constrained by Keplerian rotation, as they are all regular star-forming disc galaxies.

After matching the catalogs and applying our cuts, we retain 1015 star-forming galaxies in our sample.

\begin{figure*}
    \centering
    \includegraphics[width=2\columnwidth]{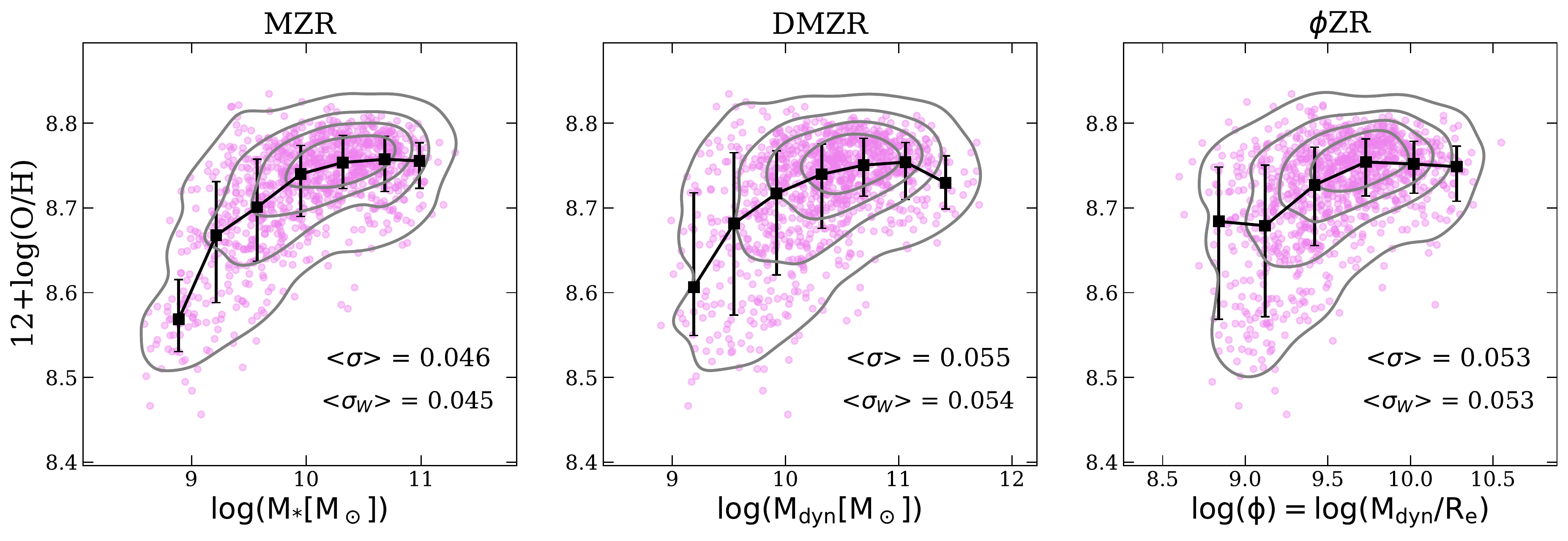}
    \caption{Gas-phase metallicity versus, stellar mass (left), dynamical mass (middle), and the gravitational potential $\phi$=$\rm M_{dyn}/R_e$ (right). The black squares are calculated in bins of stellar mass (left), dynamical mass (middle) and $\phi$ (right), where the values correspond to the bin median and the errorbars to the 16th and 84th percentiles. The density contours show the distribution of the galaxies, with the outer contour enclosing 95\% of the galaxy population. The points and contours in the figure show how the the mass-metallicity relation (MZR) in the left panel appears to be significantly less dispersed than either the dynamical mass-metallicity relation (DMZR), in the middle panel, or the $\phi$-metallicity relation ($\phi$-ZR), in the right panel. The panels also show much stronger and tighter correlation of metallicity with M$_*$ than with $\rm M_{dyn}$ or $\phi$. In the bottom-right corner of each panel we also report the average dispersion around each relation  ($\sigma$) and the average dispersion weighted by the number of galaxies in each bin ($\sigma_W$), which both quantitatively indicate that the DMZR and the $\phi$-ZR are more scattered than the MZR.
    This indicates that the MZR is not simply tracing a more fundamental dependence on dynamical mass or $\phi$. 
    }
    \label{fig:MZRvDMZR}
\end{figure*}

\section{Methods to determine the primary metallicity drivers}

\label{sec:disp_pcc_rf}

A major issue when trying to determine dependencies between inter-correlated quantities is whether an observed correlation is in fact intrinsic or it is instead an indirect by-product of other more intrinsic correlations. This is a particularly serious problem when exploring scaling relations between galaxy properties. Indeed, most galactic properties show prominent scaling relations, but many of these are not ``causal'', they are often indirect correlations resulting from more fundamental (causal) relations.
We use multiple techniques to disentangle direct, fundamental inter-dependencies, from other relations that are instead simple by-products of indirect dependences.

\subsection{Dispersion}

Our initial approach is to plot the two quantities of interest and to analyse the  dispersion of one quantity in intervals of another quantity. Specifically,
if the dependence of metallicity on stellar mass is simply a consequence of the correlation between stellar mass and dynamical mass (or $\phi$) then we would expect the dispersion in the metallicity-dynamical mass (or metallicity-$\phi$) plane to be reduced compared to the metallicity-stellar mass plane.

\subsection{Partial Correlation Coefficients}
Another key technique is the use of partial correlation coefficients \citep[PCCs,][]{Bait2017MNRAS.471.2687B,2020BluckB, Baker2022MNRAS.510.3622B}. 
These allow us to take the correlation between two quantities whilst controlling for further quantities. This gives us the 'true', intrinsic correlation between two quantities while accounting for other indirect correlations with additional quantities.

On a 2D colour plot, i.e. quantity ``y'' versus quantity ``x'' with the points or bins colour-coded by a further quantity ``z'' \citep[as in][]{2020Bluck, Piotrowska2019, Baker2022MNRAS.510.3622B, Baker_rFMR_2022arXiv221003755B, Baker_integrated_2022arXiv221110449B},
the ratios of partial correlation coefficients can be used to define an arrow which points in the direction of the greatest increasing gradient in the color-coded (z) quantity. The angle of this arrow is then given by \citep[][]{2020BluckB}
\begin{equation}
    \text{tan}(\theta)=\frac{\rho_{yz|x}}{\rho_{xz|y}}
    \label{theta}
\end{equation}
where $\theta$ is the arrow angle measured anticlockwise from the horizontal, while $\frac{\rho_{yz|x}}{\rho_{xz|y}}$ is the ratio of the partial correlation coefficients between the y axis quantity and the colour-coded (z axis) quantity, to the x axis quantity and the colour-coded (z axis) quantity. 
The arrow enables the relative dependence of the colour-coded (z axis) quantity on the x and y axis quantities to be identified visually, whilst the arrow angle enables the relative strengths of the correlations to be determined quantitatively. 

The additional benefit of partial correlation coefficients is that we can also use them to identify whether a dependence between two quantities is positive or negative. Yet, one limitation of partial correlation coefficients is that they can only properly uncover monotonic relationships.

\subsection{Random Forest Regression} 

In addition to dispersions and partial correlation coefficients we use random forest (RF) regression. 

 At its most simple, a random forest is an assortment of decision trees, each of which attempts to reduce Gini Impurity (which gives a measure of the quality of a split) in order to determine the relative importance of a set of parameters in contributing to a given variable. The sample is split into a training and test sample with an 80-20 split. The random forest is then trained on the training sample in order to build a model. This model is then checked by being applied to the (unseen) test sample. The test and train mean squared errors are then compared to check that the model is not overfitting. For more details on random forests see \citet{Bluck2022A&A...659A.160B}.
 
 The main advantage of random forest regression is that it can probe several inter-correlated parameters simultaneously, and can even find non-monotonic relationships. This enables it to explore relationships that cannot be uncovered by the partial correlation coefficients. As was found in \citet{Bluck2022A&A...659A.160B}, random forests are able to find intrinsic dependence (i.e. causality) among many inter-correlated variables. We use random forest regression to obtain relative importances for all parameters of interest in driving the metallicity.

\section{Results}

\label{sec:results}

We start by investigating the shape of the {\it stellar} mass-metallicity relation (which we will conventionally call ``MZR'' in the following) and by exploring whether we see a similar relation for the dynamical mass or the gravitational potential, i.e. a dynamical mass metallicity relation (DMZR) or a gravitational potential-metallicity relation ($\phi-ZR$). We then determine which of the parameters is more important in driving the metallicity by utilising partial correlation coefficients and random forest regression.

\subsection{Dispersion}

First we check we obtain the expected MZR in our sample of MaNGA galaxies. We then compare it against the equivalent dynamical mass-metallicity relation (DMZR) and $\phi$-metallicity relation ($\phi$-ZR), where $\rm \phi=M_{dyn}/R_e$ and is a proxy of the gravitational potential. This enables us to explore the plausibility of the gravitational well hypothesis for the existence of the MZR.

Figure \ref{fig:MZRvDMZR} shows metallicity versus stellar mass (left), dynamical mass (middle), and gravitational potential $\phi$ (right). The black squares correspond to the median values of the metallicity in bins of the x-axis quantity, and the errorbars correspond to the 16th and 84th percentiles.  The bins are centred on the median value of the quantity in the bin, and the bin widths are 0.39dex, 0.4dex and 0.32dex in the left, middle and right plot respectively.
The density contours show the distribution of galaxies, with the outer contour enclosing 95\% of the population.
The left panel shows the standard MZR, with metallicity increasing with stellar mass until it starts to plateau at $\rm \log{(M_*/M_{\odot})}>10.5$. On the contrary, in the middle panel and the right panel, the metallicity show only weak signs of a correlation with dynamical mass and $\phi$, significantly less pronounced than that of the MZR. Most importantly, the density contours and scatter of the points show a much greater dispersion of galaxies in both the dynamical mass-metallicity plane and the $\phi$-metallicity plane, compared to the stellar mass-metallicity plane.

We quantify the different scatter of the three relations by measuring the average metallicity dispersion around each of the three scaling relation (this is obtained by taking the average of the dispersions measured in bins of the x-axis quantity in each of the three panels of  Figure \ref{fig:MZRvDMZR}). For the MZR we obtain $\langle \sigma\rangle = 0.046$, while for the DMZR and the $\phi$ZR we obtain larger average dispersions of $\langle \sigma\rangle = 0.055$ and $\langle \sigma\rangle = 0.053$, respectively. 
 We also calculate a weighted dispersion, given by $\langle \sigma_W \rangle$, where average dispersion is weighted by the number of galaxies in the bin, this ensures that we are not giving undue prominence to smaller bins. We find this is also consistent with the non-weighted average dispersion. 
This result quantitatively confirms the visual assessment that the DMZR and $\phi$ZR relations are more scattered than the MZR.

The results in this section provide an initial indication that in the MZR the stellar mass is not simply behaving as a secondary indirect proxy for either the dynamical mass or $\phi$. If this was the case we would expect to see lower dispersion, and a stronger and tighter DMZR and $\phi$ZR than the MZR. Instead, the opposite is found.

\subsection{Partial Correlation Coefficients analysis}

\begin{figure*}
    \centering
    \includegraphics[width=\columnwidth]{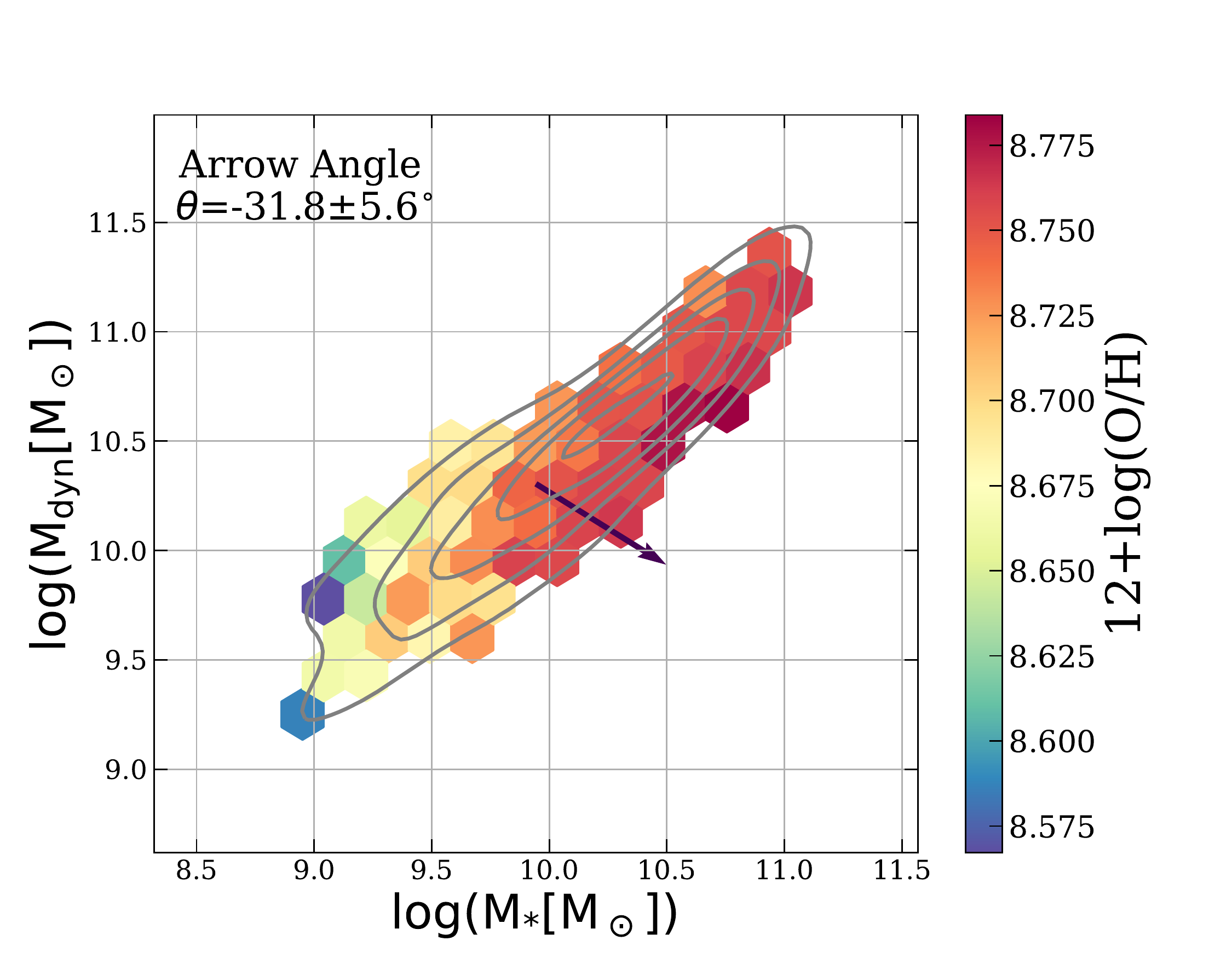}
    \includegraphics[width=\columnwidth]{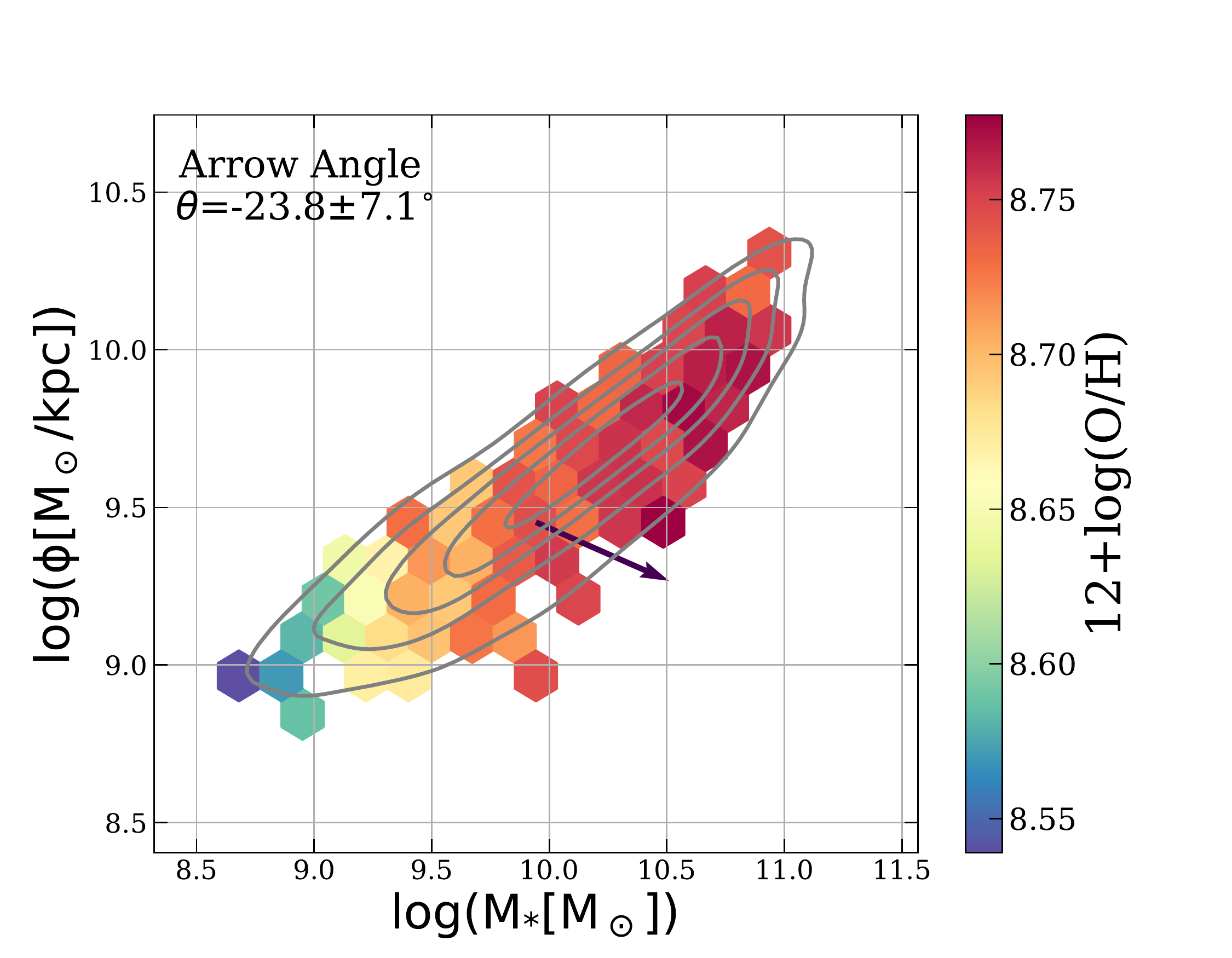}
    \caption{2D histogram of dynamical mass (left), and gravitational potential (right), versus stellar mass, where the colour-coding corresponds to the median metallicity in each bin. The density contours show the distribution of galaxies in the two panels, with the outer contour enclosing 90\% of the galaxy population. The partial correlation coefficient arrow points in the direction of the greatest increasing gradient in metallicity. The arrow shows that the metallicity is primarily driven by the stellar mass in both instances. This means that the stellar mass (i.e. MZR) is significantly more important in driving the metallicity than either the dynamical mass or the gravitational potential. Interestingly, at a fixed stellar mass, the metallicity dependence on either M$_{\rm dyn}$ or $\phi$ is inverted, i.e. in the opposite direction of what the metal retention scenario by gravity would predict.}
    \label{fig:hex_plot}
\end{figure*}

We now explore the primary driver of the metallicity by applying partial correlation coefficients to the galaxies within our sample.

Figure \ref{fig:hex_plot} shows the 2D histograms of dynamical mass (left), and $\rm \phi=M_{dyn}/R_e$, the gravitational potential (right), versus stellar mass, colour-coded by the median metallicity in each bin. The density contours show the distribution of galaxies, where the outer density contour encloses 90\% of the galaxies within the sample. 
The contours of the plot show a tight relationship between the stellar mass of the galaxies and their dynamical mass and gravitational potential respectively. Within such correlation, the PCC arrow points in the direction of the greatest increasing gradient in the metallicity. The arrow shows that, in both the left and right panels, stellar mass is significantly more important than either the dynamical mass or the gravitational potential (the arrow is almost horizontal). The arrow angle does appear to show a slight dependence on dynamical mass although we caution that in this type of diagram we are only controlling for stellar mass when taking the partial correlation coefficients, so there are potentially other quantities dependencies that could be tied up in a small correlation with dynamical mass. The random forest regression will enable us to disentangle these. In addition, the small correlation we see here is actually negative suggesting that, even if there is a small intrinsic dependence upon dynamical mass, it {\it anticorrelates} with metallicty, i.e. the greater the dynamical mass (at fixed M$_*$), the lower the metallicity. This is the complete opposite to that expected by the gravitational retention of metals scenario.

In the scenario where gravitational gas retention is responsible for the MZR one would expect the stronger correlation with the gravitational potential $\phi$. However, Figure \ref{fig:hex_plot}, and in particular the PCC arrow, shows that, at a fixed stellar mass, the metallicity is nearly independent of $\phi$ and, if anything, anti-correlates with $\phi$ (at fixed M$_*$).

\subsection{Random Forest analysis}

\begin{figure}
    \centering
    \includegraphics[width=\columnwidth]{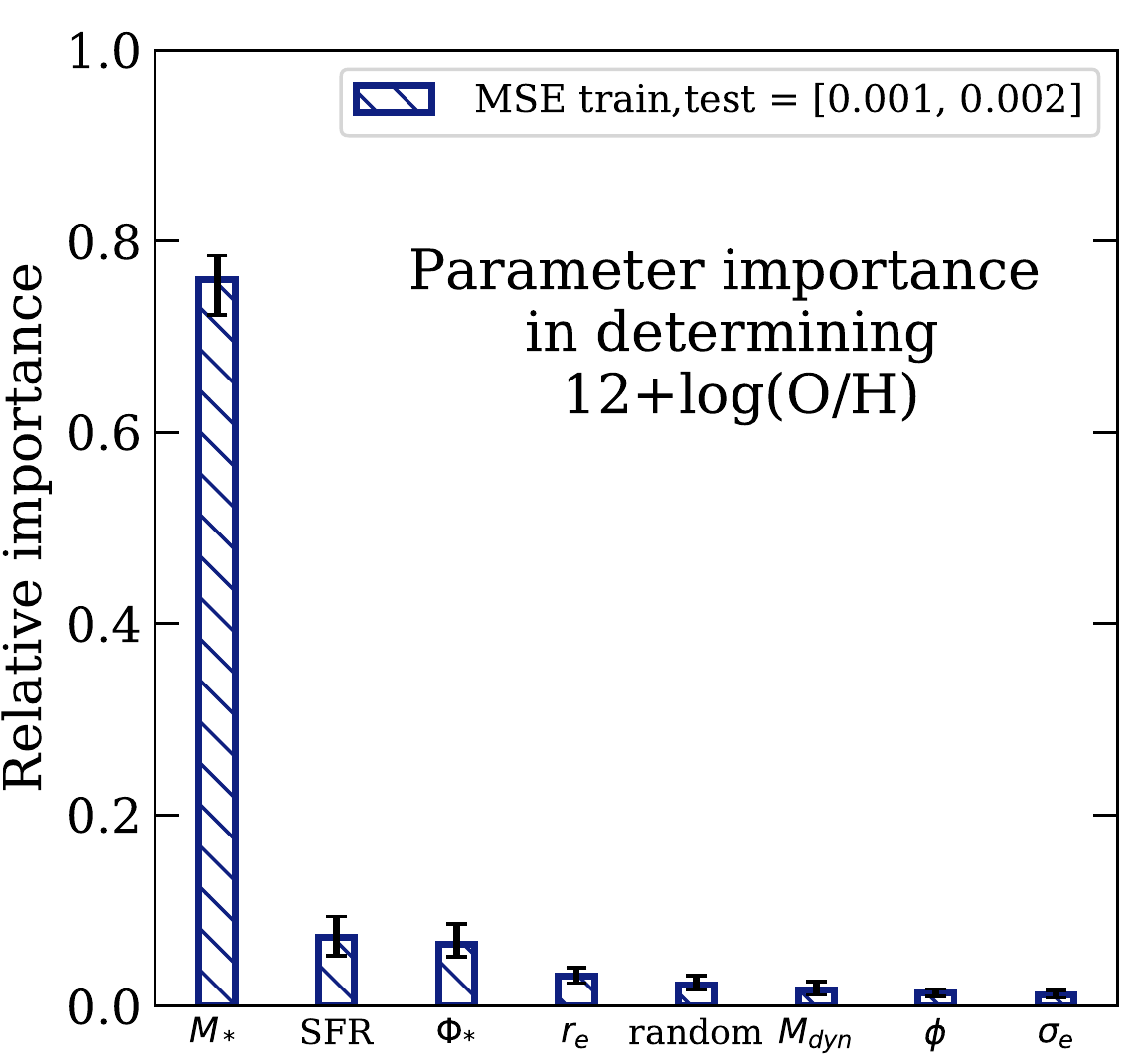}
    \caption{ Bar-chart showing the random forest regression parameter importances for determining the metallicty. The parameters evaluated are stellar mass ($M_*$), effective radius ($r_e$), dynamical mass ($\rm M_{dyn}$), star formation rate (SFR), gravitational potential ($\rm \phi=M_{dyn}/R_e$), $\rm \Phi_{*}=M_{*}/R_e$, the velocity dispersion ($\sigma_e$) and a control uniform random variable. The errors are obtained by bootstrap random sampling 100 times. The figure shows that the metallicity is driven almost entirely by the stellar mass with little to no contribution from the remaining parameters (although there is a slight contribution from SFR and  $\Phi_*$). This means that it is the stellar mass responsible for determining the metallicity, not the dynamical mass or gravitational potential.}
    \label{fig:RF}
\end{figure}

The next stage is to analyse all these parameters simultaneously by employing a random forest analysis. This enables us to probe non-monotonic trends in addition to being able to uncover the intrinsic dependencies.
In addition to the aforementioned parameters, we also explore the effect of including, the velocity dispersion ($\sigma_e$), and the effective radius ($r_e$) into our analysis. 
Following the findings of \cite{D'Eugenio_potential_2018MNRAS.479.1807D}, \cite{Vaughan_Z_phi_2022MNRAS.tmp.2242V}, \cite{Barone2020ApJ...898...62B} and \citet{Cappelari_M/R_2022arXiv220814974C} we also include in our analysis $\Phi_*$ where $\rm \Phi_{*}=M_{*}/R_e$. We note that it was not possible to test this version of $\Phi$ in the previous partial correlation coefficient analysis as, by definition, $\Phi_{*}$ contains $M_*$ and so cannot be compared to $M_*$ with PCCs. Finally, we also include the star formation rate (SFR), as the metallicity is expected to have a secondary dependence on SFR following the so-called Fundamental Metallicity Relation \citep[FMR][]{2010Mannucci}.

Figure \ref{fig:RF} is a bar-chart showing the importance of all the galactic parameters discussed above in determining the metallicity. We also include a random uniform variable as a control that it provides approximately zero importance. The errorbars are obtained by bootstrap random sampling 100 times. The bar-chart shows that the metallicity is almost entirely determined by the stellar mass with little to no contribution from the remaining parameters.

The Random Forest
 rules out dynamical mass, gravitational potential or velocity dispersion as being the main drivers of the MZR.
 This clearly indicates that the scenario of metals retention by the gravitational potential is not a viable explanation for the origin of the MZR. The strong metallicity dependence on M$_*$ is not because M$_*$ is a proxy of M$_{\rm dyn}$ or of the gravitational potential; instead, there is a direct, fundamental relationship between metallicity and stellar mass.
 
We also note that
there is some slight, secondary dependence on the SFR, which is expected from the FMR, however the importance ($<$10\%) is much lower than predicted by the original work of \cite{2010Mannucci}, which indicates a contribution of the SFR of about 30\% in determining the metallicity.
Within this context, it is interesting that the RF also finds a slight dependence on $\Phi_{*}$, with about the same importance as the SFR; therefore, we find that a residual correlation between stellar mass, metallicity and $\Phi_{*}$ (a M-Z-$\Phi_{*}$ relationship) could exist at a similar significance level to the FMR.
%therefore, it is possible that part of the importance originally associated with SFR in determining the metallicity, is actually partly to ascribe to $\Phi_{*}$.
A dependence of the metallicity on $\Phi_{*}$, was already found by \citet{D'Eugenio_potential_2018MNRAS.479.1807D} and \citet{Cappelari_M/R_2022arXiv220814974C}. They ascribed it to $\Phi_{*}$ tracing the gravitational potential of the galaxy, hence the metals retention effect; however, we have verified that this is not the case, as when we actually determine more accurately the gravitational potential as $\rm \Phi = M_{dyn}/R_e$, this does not appear to play any role in determining the metallicity (Figure \ref{fig:RF}); therefore, the role of $\Phi_{*}$ in determining the metallicity must be associated with some other physical process.
However, we also caution that, as $\Phi_{*}$ is composed of the most important parameter in the RF (M$_*$), and with our modest sample size, we cannot draw  solid conclusions on the role of $\Phi_{*}$ in determining the metallicity (aside from the obvious that $\Phi_{*}$ is nowhere near as important as stellar mass in being the primary driver of the metallicity).

\section{Discussion}

\label{sec:discussion}

\subsection{The role of uncertainties}

One may wonder whether the results might be affected by the uncertainties in the dynamical mass estimates being larger than for the stellar masses. Actually it is the other way around. In this sample of well resolved star-forming galaxies the dynamical masses are very well constrained as they are all simple, regular (dynamically) cold rotators, with neat rotation curves. Essentially their dynamical masses are pinned down down primarily by the Keplerian rotation, i.e. by one of the oldest laws in physics. Indeed \cite{Li_dynamicalMass_2018MNRAS.476.1765L} quote only a 10\% error on the $\rm M_{dyn}$ \citep[see also ][]{Cappellari_2013MNRAS.432.1709C}. On the contrary, stellar masses are much more uncertain, as they rely on stellar population models and initial mass function assumptions. The stellar masses are particularly uncertain when near-IR data is not available, as it is in this case. The uncertainties quoted for stellar masses are typically $\sim 0.5$dex. Therefore, the tighter correlation of metallicity with M$_*$ than with $\rm M_{dyn}$ cannot be explained in terms of lower putative uncertainties on M$_*$. 

\subsection{Possible scenarios}

These results present a clear challenge to the idea that it is the underlying gravitational potential being able to retain metals that explains the MZR.
Were this to be the case we would expect the dependence of metallicity on stellar mass to be replaced by that of either dynamical mass or the gravitational potential. This is not found.

Instead we propose a different, much simpler explanation for the mass-metallicity relation, which is consistent with our findings. Stellar mass is, by definition, simply the integral of star formation rate (SFR) over time. Star formation rate itself is proportional to the number of supernovae (SNe) explosions. Metallicity increases with increasing number of SNe explosions. Hence, stellar mass is proportional to the integrated SNe rate which is itself proportional to the integrated metallicity production. This means as stellar mass increases the metallicity also increases, resulting in a mass-metallicity relation. 
Mathematically, this is described in the following equation
\begin{equation}
    \rm M_*=\int SFR\, dt\, \propto \int SNR\, dt\, \propto \int \frac{dZ}{dt}\,dt \propto Z
    \label{eq:metal}
\end{equation}
where SNR is the rate of supernovae and Z is the metallicity.
This does not require any gravitational potential effects.
In summary, the MZR is simply tracing the fact that M$_*$ is proportional to the integral of metals production. Outflows can obviously be present, remove metals and their effect embedded in the proportionality constants in Eq. \ref{eq:metal}. However, in this scenario the outflow loading factor can be independent of mass and outflows do not necessarily need to be heavily metal-loaded. This approach can also reproduce the MZR well when combined with a gas-regulator model \citep[][]{Lilly2013, Peng2014FromHaloes}. However, we note that this model, whilst accounting for the low to intermediate mass-metallicity regime well, must be a simplification, as it would not be able to reproduce the MZR plateau observed at high stellar masses.

An interesting avenue of further research would be whether the gravitational potential has an effect on the local resolved metallicity of a galaxy. It could contribute more due to internal mixing of enriched and metal poor gas within the galaxy (rather than being dependent on galaxy wide inflows/outflows). \cite{Baker_rFMR_2022arXiv221003755B} found that the resolved metallicity appears to have an intrinsic dependence upon both local and global galactic quantities and that a dependence upon galactocentric radius remains even once other parameters are accounted for. This could be linked to variations in the gravitational potential influencing the ability of both local galaxy scale winds, together with other factors, such as mixing of metals via fountains or radial gas flows.

\section{Conclusions}

\label{sec:Conclusions}

We have investigated the origin of the mass-metallicity relation, i.e. the
relation between stellar mass and gas metallicity, by using spatially resolved
spectroscopy of more than 1,000 star forming galaxies from the MaNGA survey to
obtain information also on the dynamical mass and gravitational potential.

We explore and disentangle the dependence of the gas metallicity on different
galaxy parameters (which are inter-correlated) by using different statistical
methods. Specifically, we use the average dispersion, Partial Correlation
Coefficients and Random Forest regression analysis.

We obtain the following results:

\begin{itemize}

\item All methods indicate that the {\it stellar} mass is the parameter that
primarily and most fundamentally drives the gas metallicity in galaxies.

\item Once the primary dependence on stellar mass is taken into account, we find
that the gas metallicity does not show any significant dependence on dynamical
mass, gravitational potential or velocity dispersion. If anything, we find that,
at fixed stellar mass, the gas metallicity slightly {\it anti-}correlates with
dynamical mass and gravitational potential.

\item Our results imply that in the MZR the stellar mass is {\it not} an indirect proxy
of the dynamical mass or of the gravitational potential of the galaxy.

\item We find a secondary dependence of the gas metallicity on the star formation
rate, although less important than what found by previous studies.

\item We also find a secondary dependence on $\rm \Phi _*=M_*/R_e$, at a level of
importance similar to that of the SFR.

\end{itemize}

Based on these finding we infer that:

\begin{itemize}

\item Our results do {\it not} support the scenario in which the MZR
is a consequence of the metals being more
effectively retained by the deeper gravitational potential of
more massive galaxies.

\item A much simpler explanation of the MZR is that the stellar mass directly
(and fundamentally) correlates with metallicity as it is proportional to the integral
of metals production.

\item We suggest that part of the correlation between SFR and metallicity found
in previous studies is resulting from the metallicity correlation between $\Phi
_*$ and metallicity. The latter parameter has an importance similar to SFR in driving the
gas metallicity, although the physical origin of such correlation remains to be explored.

\end{itemize}

\section*{Acknowledgements}
We are very grateful for constructive feedback and comments from Francesco Belfiore, Asa Bluck, Michele Cappellari, Mirko Curti, Francesco D'Eugenio and Piyush Sharda.
We thank the referee for their constructive comments which have helped this paper. 

W.B., and R.M. acknowledge support by the Science and Technology Facilities Council (STFC) and ERC Advanced Grant 695671 "QUENCH". RM also acknowledges funding from a research professorship from the Royal Society.

Funding for the Sloan Digital Sky Survey IV has been provided by the Alfred P. Sloan Foundation, the U.S. Department of Energy Office of Science, and the Participating Institutions. SDSS acknowledges support and resources from the Center for High-Performance Computing at the University of Utah. The SDSS web site is www.sdss.org.

SDSS is managed by the Astrophysical Research Consortium for the Participating Institutions of the SDSS Collaboration including the Brazilian Participation Group, the Carnegie Institution for Science, Carnegie Mellon University, Center for Astrophysics | Harvard \& Smithsonian (CfA), the Chilean Participation Group, the French Participation Group, Instituto de Astrofísica de Canarias, The Johns Hopkins University, Kavli Institute for the Physics and Mathematics of the Universe (IPMU) / University of Tokyo, the Korean Participation Group, Lawrence Berkeley National Laboratory, Leibniz Institut für Astrophysik Potsdam (AIP), Max-Planck-Institut für Astronomie (MPIA Heidelberg), Max-Planck-Institut für Astrophysik (MPA Garching), Max-Planck-Institut für Extraterrestrische Physik (MPE), National Astronomical Observatories of China, New Mexico State University, New York University, University of Notre Dame, Observatório Nacional / MCTI, The Ohio State University, Pennsylvania State University, Shanghai Astronomical Observatory, United Kingdom Participation Group, Universidad Nacional Autónoma de México, University of Arizona, University of Colorado Boulder, University of Oxford, University of Portsmouth, University of Utah, University of Virginia, University of Washington, University of Wisconsin, Vanderbilt University, and Yale University.

%The Acknowledgements section is not numbered. Here you can thank helpful
%colleagues, acknowledge funding agencies, telescopes and facilities used etc.
%Try to keep it short.

%%%%%%%%%%%%%%%%%%%%%%%%%%%%%%%%%%%%%%%%%%%%%%%%%%
\section*{Data Availability}

The MaNGA data that is used in this work is publicly available at 
https://www.sdss.org/dr15/manga/manga-data/.
The dynamical masses, effective radii and velocity dispersions are available from \citet{Li_dynamicalMass_2018MNRAS.476.1765L}.

%%%%%%%%%%%%%%%%%%%% REFERENCES %%%%%%%%%%%%%%%%%%

% The best way to enter references is to use BibTeX:

\bibliographystyle{mnras}
\bibliography{example} % if your bibtex file is called example.bib

\appendix

%%%%%%%%%%%%%%%%%%%%%%%%%%%%%%%%%%%%%%%%%%%%%%%%%%

% Don't change these lines
\bsp	% typesetting comment
\label{lastpage}
\end{document}